\documentclass{ws-p8-50x6-00}

\begin{document}

\title{Colour Reconnection at LEP2}

\author{P. Abreu}

\address{LIP and IST,\\
Av. Elias Garcia, 14, 1st, 1000-149 Lisboa, Portugal\\ 
E-mail: abreu@lip.pt}


\maketitle

\abstracts{
The preliminary results on the search of colour reconnection effects (CR)
from the four experiments at LEP, Aleph, Delphi, L3 and Opal, are
reviewed.
Extreme models are excluded by studies of standard variables, and
on going studies of a method first suggested by L3, the particle
flow method\cite{L3pf}, 
%to search for Colour Reconnection in the scope of
%more realistic models, 
are yet inconclusive.}

\section{Introduction}

Colour reconnection, better said colour rearrangement between partons
or cross-talk (CR),
corresponds to a colour interference between partons close-by in space-time, and
is a good probe to understand the dynamics of hadronization.
It is expected to be the cause for the J/$\psi$ formation in the decays of the
B meson, in which occurs a cross-talk between two original colour singlets,
$\bar{\mathrm c}$+s and c+spectator\cite{sk98}.

In the case of double production of heavy particles
(WW, ZZ, ZH, t$\bar{\mathrm t}$), if 
both decay hadronically almost simultaneously in space and time, 
there could be also cross-talk effects between the decay products.
In this case
particles cannot anymore be assigned unambiguously to one parent particle,
and the parent particle's properties such as invariant mass cannot be inferred
from its alleged decay products. 
Finally it could also induce additional interference between beam and final
state partons in hadron machines.

LEP2, the second phase of the LEP machine at CERN, has been working since
1996 at centre
of mass energies above the threshold for double W boson production, aiming at
measuring the W boson mass with a precision of the
order of 50 MeV~\cite{lep2mW}. 
Each of the four LEP experiments collected about 10000 WW events,
for a total luminosity per experiment of about 700 pb$^{-1}$.

In the hadronic channel, in which both W bosons decay hadronically, the
decay products from different W bosons can interfere, since the distance
traveled by the W boson before decaying, 
$c\tau_W\approx{\hbar}c/\Gamma_W\approx 0.1$fm, is much smaller than the
typical hadronization scales of 1~fm. At several stages these effects could
be caused by 
colour rearrangement between the quarks coming from the W bosons,
gluon exchange in the parton cascade, and by
Bose Einstein interference between identical bosons (e.g, pions).
The third item is the topic of other proceedings in this conference
(see Jorn van Dalen's contribution), and
the first two items are the main subject of the review presented below,
with most of the results preliminary.

\section{Theoretical aspects of colour rearrangements}

The effects of colour rearrangement between the primary quarks 
or energetic hard gluons from different W bosons, are small at perturbative
level, with the
effect in the W mass of the order of 

\begin{equation}
(\delta M_W)_{PT}\approx \left(\frac{C_F\alpha_s(\Gamma_W)}{\pi}\right)^2
                               \frac{1}{N_C^2-1}\Gamma_W 
                 \approx O(1\mathrm{~MeV})\,.
\label{qcdPT}
\end{equation}

Non perturbative QCD effects between soft gluons ($E_g<\Gamma_W$) 
coexisting in space and time
could be large, of $O(10\mathrm{~MeV})$ in the W
mass, and affect average multiplicities and inclusive particle
distributions, 
with enhancements for low-momentum and/or heavier particles. To estimate these
effects phenomenological models are needed and were developed in the past
by T. Sj\"ostrand and V. A. Khoze \cite{sk94} (skI, skII, skII'), 
G. Gustafson and J. Hakkinen \cite{gh} (gh), L. L\"onnblad \cite{lonnblad}
(ar2, ar3), G. Marchesini {\it et al} \cite{hw} (herwig), J. Ellis and K. Geiger
\cite{eg} (eg), and more recently by J. Rathsman \cite{rathsman}. 

The models of Sj\"ostrand and Khoze, implemented in the Pythia Monte Carlo
generator~\cite{pythia}, type I and type II in analogy to the super-conducting
vortices associated to the strings, allow for reconnection if the strings
cross each other. 
%In type II model (skII), the string is a vortex line,
%and when two lines cross they reconnect with unit probability
%(model skII' only allows for reconnections that diminish the total string
%length). 
In type I model (skI), the most commonly used as a basis for the
studies reported here, the strings have a transverse dimension, similar to
flux tubes, and they may cross with different overlapping volumes
$V_{\mathrm{overlap}}$. Then the
probability of reconnection in one event is given by 
$P_{\mathrm{reco}} = 1 - e^{-k_I V_{\mathrm{overlap}}}$,
with $k_I$ a user parameter allowing to vary the percentage of reconnected
events (probability of reconnection integrated for the accepted phase space).

\section{Analysis of standard variables}

The effects on the average charged multiplicities $n(4q)$ in fully hadronic WW
events (WW(4q)), were the first signatures probed by the four experiments, in
particular 
by comparing with the average charged hadronic 
multiplicities $n(2q)$ in semi-leptonic WW events 
(see~\cite{delphiww1997,opalww1997} and~\cite{deangelis1997}
for a review).
In the absence of these effects, there should not be any
 statistically significant
difference between $n(4q)$ and twice $n(2q)$. The latest
results~\cite{alephww,delphiww2000,l3ww,opalww} on the average
multiplicities in WW events and on their differences and ratios 
are shown in table~\ref{tab:mult}. The Aleph result was not included in the
averages because is not corrected for momentum acceptance and selection
biases.

\begin{table}[t]
\caption{Average charged multiplicities in WW(4q) ($n(4q)$) and 
WW(2q) ($n(2q)$) events, their differences and ratios. The Aleph
results are not corrected for momentum acceptance and selection biases.
\label{tab:mult}}
\begin{center}
\footnotesize
%\begin{tabular}{|c|c|c|c|c|c|}
\begin{tabular}{|c|c|c|c|c|}
\hline
Exp. & % $\sqrt{s}$ (GeV)       &
 $n(4q)$                 & 
 $n(2q)$                 & $n(4q)-2n(2q)$         & n(4q)/(2n(2q)) \\
\hline
Aleph      & % 183-202          &
 35.75$\pm$0.13$\pm$0.52 &
 17.41$\pm$0.12$\pm$0.15 & 0.93$\pm$0.27$\pm$0.29 & 1.027$\pm$0.008$\pm$0.007\\
Delphi     & % 183-189          &
 38.87$\pm$0.29$\pm$0.29 &
 19.57$\pm$0.26$\pm$0.23 &-0.32$\pm$0.60$\pm$0.54 & 0.990$\pm$0.015$\pm$0.011\\
L3         & % 183-202          &
 37.90$\pm$0.14$\pm$0.41 &
 19.09$\pm$0.11$\pm$0.21 &-0.29$\pm$0.26$\pm$0.30 & 0.993$\pm$0.007$\pm$0.015\\
Opal       & % 183-189          &
 38.51$\pm$0.22$\pm$0.33 &
 19.25$\pm$0.16$\pm$0.16 & 0.06$\pm$0.39$\pm$0.32 & 1.003$\pm$0.010$\pm$0.012\\
\hline
\multicolumn{3}{|c|}{\hfill Average(DLO)} 
                         &-0.18$\pm$0.21$\pm$0.21 & 0.996$\pm$0.006$\pm$0.008\\
\hline
\end{tabular}
\end{center}
\end{table}

It can be concluded that the measurements are compatible with equity
between WW(4q) and twice WW(2q), and do not show any evidence for
colour rearrangement effects. 

These effects were also searched for
in inclusive distributions of charged particles,
as well as in the ratio of the average multiplicities of low momentum 
identified heavy hadrons (Kaons and protons) in WW(4q) to 
WW(2q)~\cite{delphiww2000,opalkp}, and no effects were found.

%\begin{eqnarray*}
%\mathrm{K~}(0.2<p<1.25\mathrm{~GeV/c}): & 0.96\pm0.38\pm0.08 &
%\mathrm{(Delphi)} \\
%\mathrm{p~}(0.2<p<1.25\mathrm{~GeV/c}) : & 0.72\pm0.57\pm0.08 &
%\mathrm{(Delphi)} \\
%\mathrm{K+p~} (0.002<x_p<0.012):         & 1.05\pm0.06\pm0.05 &
%\mathrm{(Opal)}\,.\\
%\end{eqnarray*}
%
%Delphi has also studied oriented event distributions, in particular the
%momentum out of the W planes, and no effect was seen
%(figure~\ref{fig:ptout2j}).
%
%\begin{figure}[t]
%\begin{center}
%%\figurebox{20pc}{15pc}{} % to have a box alone
%\epsfxsize=15pc \epsfbox{delphi_ptout2j_plt2.ps} % postscript image file name
%\caption{$P_{Tout}$ of the W decaying dijet plane for soft charged particles,
% WW(4q)-2W(2q) data compared to simulation with no CR.
%\label{fig:ptout2j}}
%\end{center}
%\end{figure}

The conclusions of the studies on the standard variables are that the extreme
models (eg, ar3) can be excluded
by these data~\cite{opalww}, but these variables are insensitive to
most of  
the other more realistic models.

\section{The particle flow method}

L3 has developed a new approach~\cite{L3pf,L3pf01}, toward a more
restrictive event selection criteria, and building new variables relating
the particle and energy distributions with respect to jets.
The tight event selection criteria
enable the proper definition of inside W and outside W regions,
but with the disadvantage of very low efficiency (less than $\approx$15\%).
This event selection has been followed also by the Delphi
collaboration~\cite{delphipf} as the mainstream analysis, and by
the other collaborations as a cross-check analysis
(Aleph~\cite{alephpf}, Opal~\cite{opalpf});
table~\ref{tab:pflum} gives the luminosities, numbers of events analysed
so far,
the efficiency of the event selection(s), the purities of the selected
data samples, and the efficiency for correct pairing of jets to their
parent W bosons 
(all the results are preliminary).

\begin{table}[t]
\caption{Luminosities, numbers of selected events used in the 
analysis, its efficiency, purity of the selected samples,
and efficiency for correct pairing of jets to the parent W.
AlephXC and OpalXC stand for the cross check analysis of Aleph and Opal.
\label{tab:pflum}}
\begin{center}
\footnotesize
\begin{tabular}{|c|c|c|c|c|c|c|c|}
\hline
 Exp.  & $\sqrt{s}$ (GeV) &{$\cal L$}(pb$^{-1}$) & Data & Expected &
   Eff.    &  Pur. & Pair \\
\hline
Aleph  & 189-208          & 626.4                & 5487 & -        &
0.92-0.88 & 0.78      & 0.7      \\
AlephXC& 189-208          & 626.4                & 684  & -        &
0.15-0.09 & 0.85-0.82 & 0.9-0.85 \\
Delphi & 183-208          & 601.4                & 759  & 721.9    &
0.15-0.09 & 0.83-0.77 & 0.76     \\
L3     & 189-208          & 626.6                & 666  & 689.9    &
0.14-0.09 & 0.85      & 0.93-0.88\\
Opal   &   189            & 182.5                & 699  & -        &
0.42      & 0.83      & 0.5      \\
OpalXC &   189            & 182.5                & 260  & -        &
0.16   & -            & -        \\
\hline
\end{tabular}
\end{center}
\end{table}

\begin{figure}[t]
\begin{center}
%\figurebox{20pc}{15pc}{} % to have a box alone
\epsfxsize=10pc \epsfbox{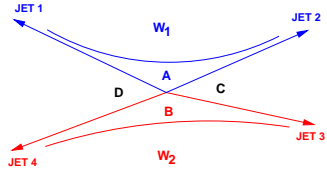}
\caption{The definition of inside W and outside W regions.
\label{fig:order}}
\end{center}
\end{figure}

For the events selected, which contain 4 jets,
distributions of the particle and energy flows
are built in a way to best reflect the inside W and outside W regions,
depicted as regions A,B (inside W) and C,D (outside W) in 
figure~\ref{fig:order}.

\begin{figure}[t]
\begin{center}
%\figurebox{20pc}{15pc}{} % to have a box alone
\epsfxsize=15pc \epsfbox{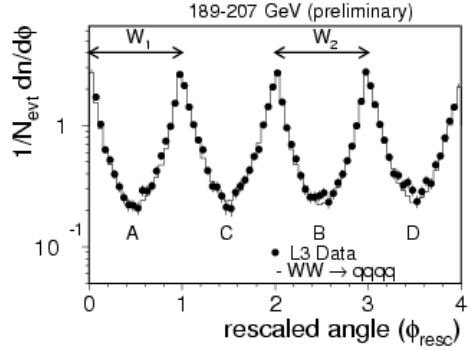}
\caption{The particle flow distribution of L3.
\label{fig:pflow}}
\end{center}
\end{figure}

The method used to build these distributions, 
explained in detail 
in~\cite{delphipf}, uses the angles between
particles and jets, and angles between jets, to define a rescaled angle
and associate the particles
to inside W and outside W regions. The rescaled angle distribution 
for the particle flow of L3 is shown
in figure~\ref{fig:pflow}.

\begin{figure}[t]
\begin{center}
%\figurebox{20pc}{15pc}{} % to have a box alone
\begin{tabular}{cc}
\epsfxsize=12pc \epsfysize=10pc \epsfbox{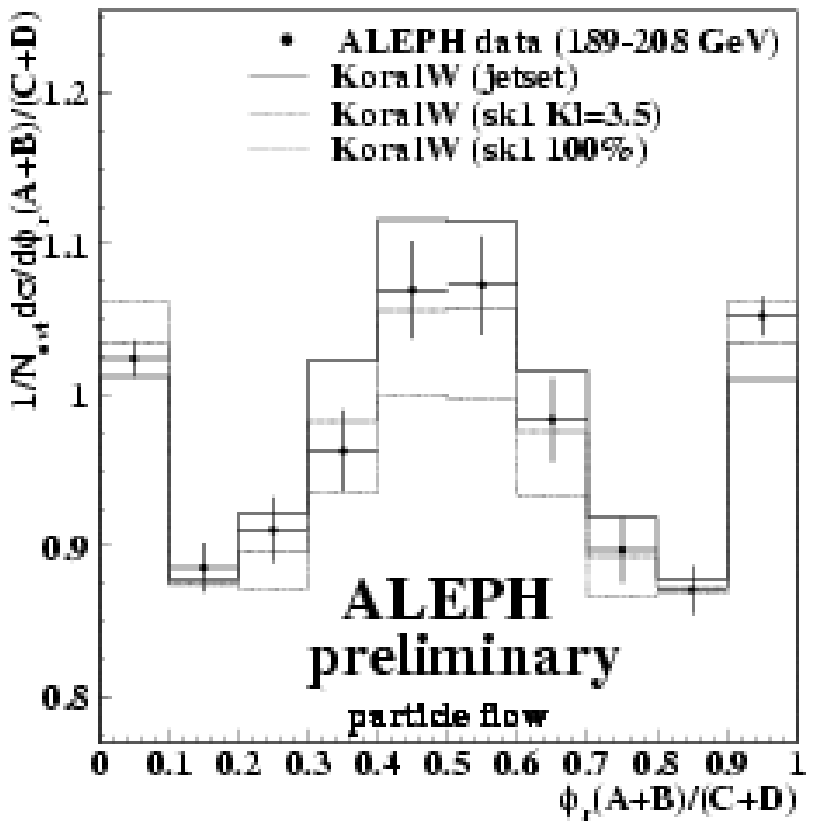} &
\epsfxsize=12pc \epsfbox{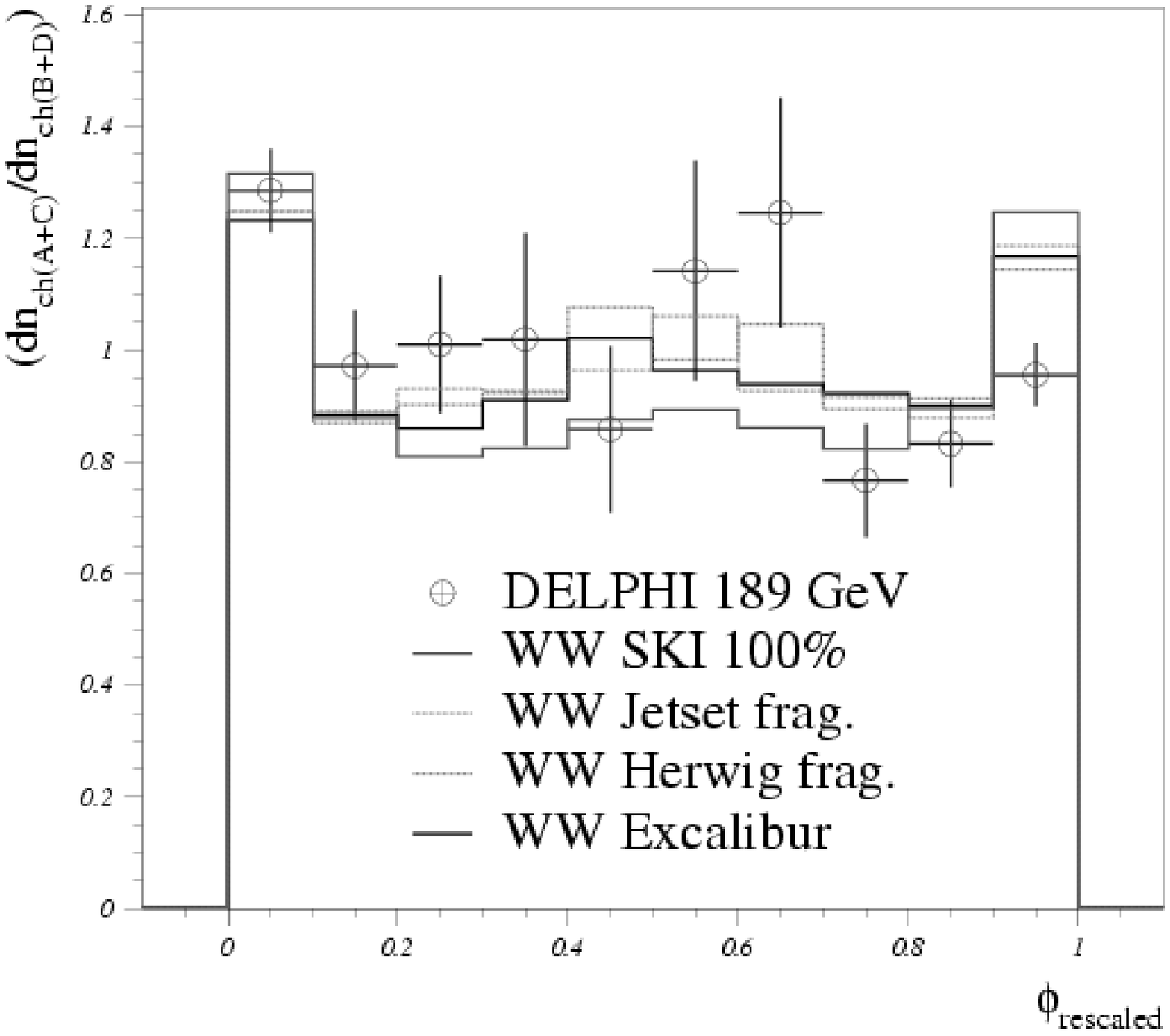}  \\
\epsfxsize=12pc \epsfbox{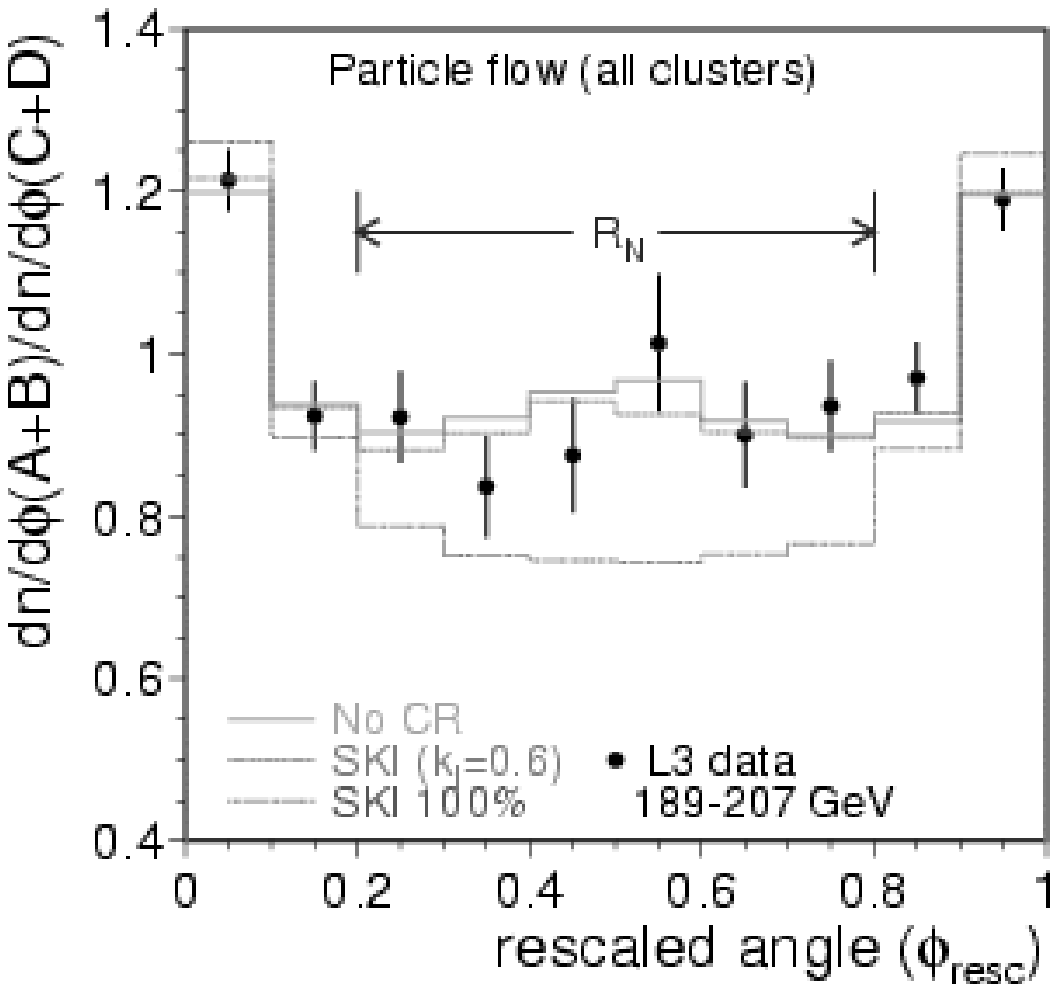}          &
\epsfxsize=12pc \epsfysize=10pc \epsfbox{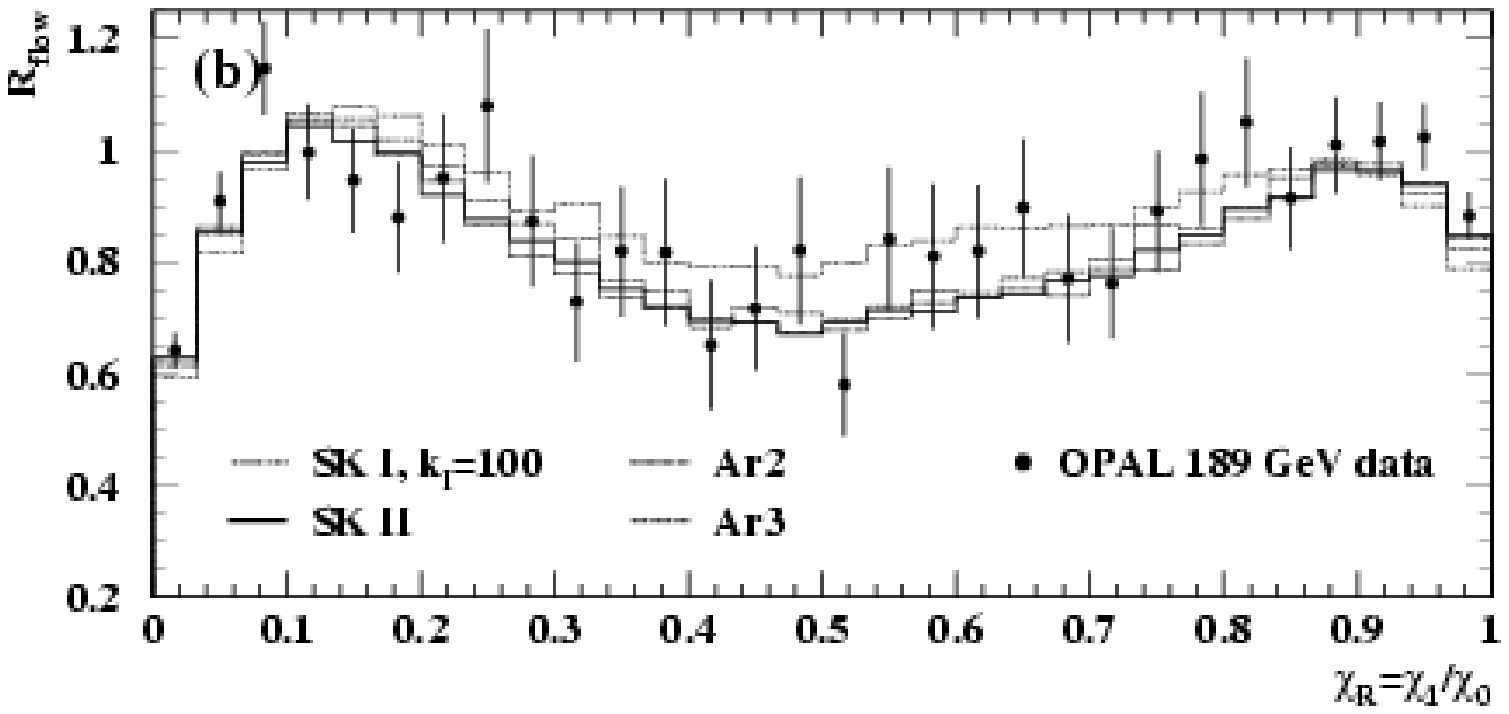}      \\
\end{tabular}
\caption{The distribution of the ratios of inside W to outside W regions,
in bins of rescaled angle (particle flow), for Aleph, Delphi, L3 and Opal.
\label{fig:ratio}}
\end{center}
\end{figure}

Adding the inside W regions and the outside W regions, the distributions 
of the ratios of inside to
outside are shown in figure~\ref{fig:ratio} for the four experiments. 

The ratios R of the integrals, from 0.2 to
0.8 in the rescaled angle, of the particle flow distributions of inside W
regions to outside W 
regions (sum of inside divided by sum of outside), are shown in 
table~\ref{tab:ratio} for the four 
experiments~\cite{alephpf,delphipf,L3pf,opalpf}, along with the
values expected from simulation with and without CR effects. The
experiment's sensitivity to the model skI with 100\% reconnection probability,
computed as  
$S=|R_{\mathrm{noCR}}^{MC}-R_{\mathrm{skI}}^{MC}|/
{\mathrm{Error}_{\mathrm{data}}}$ is also shown in the table. 
\begin{table}[t]
\caption{R values for the data, simulation without and with CR effects
(from model skI with 100\% reconnection probability), 
and sensitivity to the CR effects. 
OpalXC stand for the cross check analysis of Opal; the Aleph cross-check
analysis gives results compatible with the standard analysis's results shown 
here. In the data values, the first error is statistical and the second error
is systematic. 
\label{tab:ratio}}
\begin{center}
\footnotesize
\begin{tabular}{|c|c|c|c|c|}
\hline
% Exp. &  R                        &    R            &    R            & $\sigma$ \\
 R     & data                      & MC no CR        & MC (skI)        & $S$ \\
\hline
Aleph  & 1.117$\pm$0.014$\pm$0.009 & 1.164$\pm$0.002 & 1.074$\pm$0.002 & 5.5$\sigma$ \\
Delphi & 0.951$\pm$0.028$\pm$0.022 & 0.950$\pm$0.010 & 0.864$\pm$0.010 & 2.8$\sigma$ \\
L3     & 0.911$\pm$0.023$\pm$0.021 & 0.920$\pm$0.003 & 0.763$\pm$0.003 & 5.0$\sigma$ \\
Opal   & 1.205$\pm$0.044$\pm$0.015 & 1.330$\pm$0.004 & 1.147$\pm$0.004 & 3.9$\sigma$ \\
OpalXC & 1.020$\pm$0.052$\pm$0.010 & 1.025$\pm$0.005 & 0.855$\pm$0.005 & 3.2$\sigma$ \\
\hline
\end{tabular}
\end{center}
\end{table}
The R value given for Delphi is the average of the R values at
each centre of mass energy, after rescaling using simulation to the
luminosity weighted average centre of mass energy of 196 GeV.
Please note that the R values shown, uncorrected for detector effects,
cannot be directly compared between different experiments.
In the systematic errors, it were considered the effects of background
subtraction and modeling (Aleph(A), Delphi(D), L3(L) and Opal(O)), 
the Bose-Einstein effects (ADL), fragmentation modeling (ADL),
generators and tuning (DO) and definition of particles and cluster objects
(L).

L3 has performed recently~\cite{L3pf} further studies, using a sample 
of WW semi-leptonic decay events, which by definition have no 
colour rearrangement effects between the decay products of different W bosons.
L3 measures in this sample a ratio of values of R in data to R in simulation 
 of $R_R=1.011\pm0.035$, perfectly compatible with unity.

The ratio between the values of R for
the data and for simulation with or without CR effects (as expected from
skI model with 100\% reconnection probability), averaged over the centre of
mass energies for each experiment, using as weights the sum in quadrature
of the statistical and systematic errors, is shown in 
table~\ref{tab:rr}. The values for Aleph, L3 and Opal, were
estimated from the respective values of R for data and simulation for
the different centre of mass energies (the statistical error on the MC samples
was considered as systematic error). Correlations in the systematic error
were taken into account inside each experiment (correlations between
experiments were not considered important at this level).

\begin{table}[t]
\caption{Average over the centre of mass energies of the ratio of the values
of R for the data to the values of R for simulation without CR effects,
and with CR effects from skI(100\%) 
($^\dagger$~stands for the Opal/Cross check analysis, and the lines
below Aleph, Delphi and L3 values are the errors).
The first error is statistical and the second error
is systematic, and in the last line are given the distances from one in units
of standard deviation.
\label{tab:rr}}
\begin{center}
\footnotesize
\begin{tabular}{|c|c|c|c|c|}
\hline
% Exp. &  R                        &    R            &    R            & $\sigma$ \\
 $\langle R_R \rangle$ & Aleph  & Delphi & L3 & Opal \\
\hline
D/noCR  & 0.961 & 1.009
        & 0.990 & 0.906$\pm$0.033$\pm$0.011 \\
        & $\pm$0.012$\pm$0.007 & $\pm$0.030$\pm$0.019 & $\pm$0.025$\pm$0.023
        & $^\dagger$0.996$\pm$0.051$\pm$0.011 \\
        & -2.8$\sigma$              & -0.3$\sigma$ 
        & -0.3$\sigma$              & -2.7$\sigma$ / $^\dagger$-0.1$\sigma$ \\ \hline
D/CR    & 1.041 & 1.110
        & 1.194 & 1.050$\pm$0.038$\pm$0.013 \\
        & $\pm$0.013$\pm$0.008 & $\pm$0.033$\pm$0.029 & $\pm$0.039$\pm$0.028 
        & $^\dagger$1.193$\pm$0.061$\pm$0.014 \\
        & +2.7$\sigma$              & +2.5$\sigma$
        & +4.7$\sigma$              & +1.2$\sigma$ / $^\dagger$+3.1$\sigma$ \\
\hline
\end{tabular}
\end{center}
\end{table}

L3 and Delphi agree with no effect observed in their data,
Aleph results are between the model without CR effects and the model of
skI with 100\% reconnection probabiblity, and Opal has two 
analysis of similar sensitivity with incompatible preliminary results.
Using simulation of skI model for different values of reconnection
probability,
as translated in the user set $k_I$ parameter, Aleph and L3 translate their
results into limits for this parameter of $k_I<25$ (68\% Confidence Level),
with the minimum of discrepancy ($\chi^2$) at $k_I=3.5$, for Aleph, and
$k_I<1.55$(68\% Confidence Level), with the minimum of discrepancy
at $k_I=0.32$ for L3.

\section{Conclusions}

After five years of very successful LEP runs, 10,000 WW events have been
collected by each of the four LEP experiments (Aleph, Delphi, L3, Opal).

The search for colour rearrangement effects in the WW fully hadronic
events, using standard variables like average charged multiplicities and
inclusive distributions, has excluded only the most extreme models of
colour reconnection (Ellis and Geiger 'eg' and Ariadne 3 'ar3'). 
The sensitivity to other more realistic models
has been shown to be negligible in these variables.

A preliminary search following a L3 idea, the particle flow, has proven
sensitive for the different experiments, but inconclusive, as one experiment
claims part of the effect, two experiments claim no observation, and one
experiment has incompatible results in two nearly equal sensitive analysis.
However, systematic studies are still in a very preliminary stage of study,
and the exploration of the data to its full extent might improve
the results in the near future.

In order to pin down models of colour rearrangement and estimate parameters,
it is mandatory to combine the results from the four experiments in order
to reduce sizeably the statistical errors.

\section*{Acknowledgments}

I am grateful for the invitation from the organizing committee to come
to this conference, in particular to Gosta Gustafson, and for the very kind
atmosphere they were able to provide {\it in situ}. I am also grateful
to M\'ario Pimenta and Alessandro De Angelis for the help in preparing this
talk and manuscript, and to Thomas Ziegler/Aleph, Nuno Anjos/Delphi,
Dominique Duchesneau/L3 and to Nigel Watson/Opal for providing results and
fruitful discussions.

%%%%%%%%%%%%%%%%%%%%%%%% END OF DOCUMENT %%%%%%%%%%%%%%%%%%%%%%%%%%%%%%%%%%%%
\end{document}